\documentclass[%
reprint,
superscriptaddress,
 nobibnotes,
 amsmath,amssymb,
 aps,
 prl,
 floatfix,
 nobalancelastpage
]{revtex4-2}

\usepackage{graphicx}% Include figure files
\usepackage{grffile}% complicated graphic extensions
\usepackage{dcolumn}% Align table columns on decimal point
\usepackage{bm}% bold math
\usepackage[T1]{fontenc}
\usepackage[utf8]{inputenc}
\usepackage{subfigure}
\usepackage{hyperref}% add hypertext capabilities
\usepackage{xcolor}
\usepackage{soul}
\usepackage[english]{babel}
\usepackage{soul} % \st{Hellow world} to strike out text

\usepackage{xspace}

\DeclareMathOperator{\sgn}{sgn}

\newcounter{maintextfigures}

\begin{document}

%\preprint{APS/123-QED}

\title{
Experimental Tuning of Transport Regimes\\in Hyperuniform Disordered Photonic Materials
}% Force line breaks with \\
%\thanks{A footnote to the article title}%

\author{Geoffroy J. \surname{Aubry}}%
\email{geoffroy.aubry@unifr.ch}
\affiliation{%
 Département de Physique, Université de Fribourg, Switzerland
}%
\affiliation{%
 Institut de Physique de Nice, Université Côte d'Azur/CNRS, France
}%
\author{Luis S. \surname{Froufe-Pérez}}%
\affiliation{%
 Département de Physique, Université de Fribourg, Switzerland
}%
\author{Ulrich \surname{Kuhl}}%
\affiliation{%
 Institut de Physique de Nice, Université Côte d'Azur/CNRS, France
}%
\author{Olivier \surname{Legrand}}%
\affiliation{%
 Institut de Physique de Nice, Université Côte d'Azur/CNRS, France
}%
\author{Frank \surname{Scheffold}}%
\affiliation{%
 Département de Physique, Université de Fribourg, Switzerland
}%
\author{Fabrice \surname{Mortessagne}}%
\affiliation{%
 Institut de Physique de Nice, Université Côte d'Azur/CNRS, France
}%

\date{\today}% It is always \today, today,
             %  but any date may be explicitly specified

\begin{abstract}% no more than 600 characters, including spaces
We present wave transport experiments in hyperuniform disordered arrays of cylinders with high dielectric permittivity.
Using microwaves, we show that the same material can display transparency, photon diffusion, Anderson localization, or a full band gap, depending on the frequency $\nu$ of the electromagnetic wave.
Interestingly, we find a second weaker band gap, which appears to be related to the second peak of the structure factor.
Our results emphasize the importance of spatial correlations on different length scales for the formation of photonic band gaps.
\end{abstract}

\maketitle

In analogy to electronic semiconductors, dielectric materials in a periodic~\cite{Yablonovitch1987,John1987,Joannopoulos2008,Vynck2009}, quasiperiodic~\cite{Zoorob2000}, or amorphous configuration~\cite{Jin2001,Florescu2009,Liew2011,Froufe-Perez2016,Froufe-Perez2017} can all display full band gaps.
For the latter materials, due to the absence of long range order, the band gap has been associated with local resonances of the scatterers or correlated scattering clusters, which is reminiscent of the tight-binding model in electronic semiconductors~\cite{Yang2010}. In contrast to electrons, however, there exist no bound photon states making this analogy questionable. Other proposals have linked the opening of a gap directly to the suppression of density fluctuations on large length scales, known as stealthy hyperuniformity (SHU)~\cite{Florescu2009}. While the precise origin of a band gap in an amorphous dielectric material is yet unknown, the transport properties inside the gap are well understood ~\cite{Joannopoulos2008,Marichy2016,Froufe-Perez2016,Froufe-Perez2017}. In both periodic and nonperiodic band gap materials, an incident light wave enters by a finite distance $L_\mathrm{B}$, called the Bragg length, and is then totally reflected. For a slab of thickness $L$, the wave can tunnel through the material with a probability $T\sim e^{-L/L_\mathrm{B}}$.
However, outside the gap, the transport properties differ strongly. Photonic crystals either reflect, diffract into Bragg peaks, or they are transparent, which is a direct consequence of long-range order and the corresponding sharp Bragg maxima in the structure factor $S(\vec{k})$.
The situation is entirely different for amorphous materials, which scatter light strongly over a broad range of $\vec{k}$.
Recent numerical work has revealed that this leads to a rich transport phase diagram for amorphous band gap materials---with regions of transparency, Anderson localization, and light diffusion---not present in ordered materials~\cite{Froufe-Perez2017}.
In contrast to disordered photonic crystals, discussed for example in the celebrated article by Sajeev John in 1987~\cite{John1987}, the diffuse scattering and localization observed outside the gap is not a consequence of imperfections, but an inherent feature of the amorphous material~\cite{Froufe-Perez2016}.
Introduced in 2004, stealthy hyperuniformity provides an elegant way to construct such idealized disordered materials with finely tunable correlations encoded by the degree of stealthiness  $\chi$, ranging from $0\to 0.5$ before the onset of crystallization~\cite{Torquato2003,*Uche2004}. 

\begin{figure}
    \centering
    \vspace{0.3cm}
    \includegraphics[width=\columnwidth]{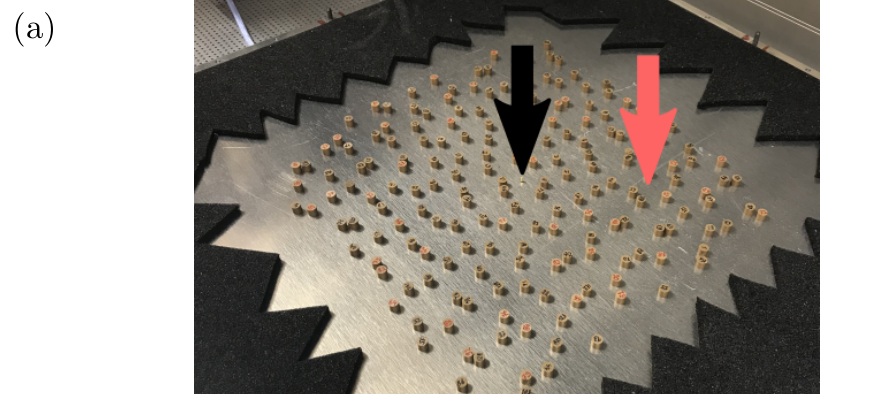}
    \includegraphics[width=\columnwidth]{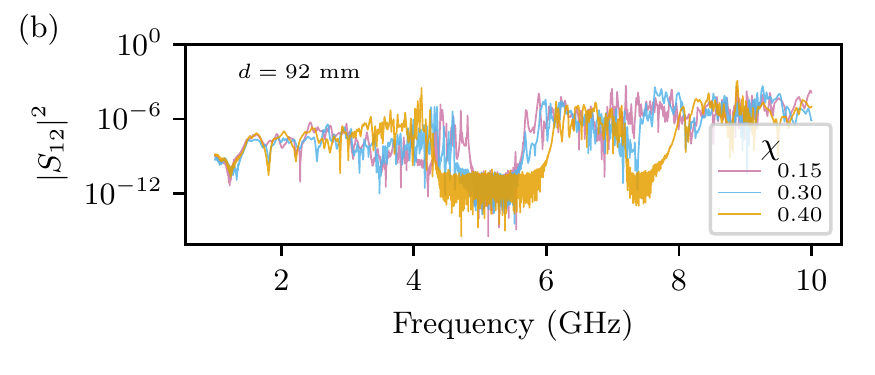}
    \caption{(a) Setup for 2D microwave scattering and transport experiments. The dielectric cylinders are placed in between two conducting aluminum plates. To reveal the interior of the sample the top plate has been removed. We place absorbing foam (LS-14 from Emerson\&Cuming) around the sample. A fixed antenna (1, black arrow) is positioned at the center of the cavity, $(x,y)=(0,0)$. The mobile antenna (2, red arrow) enters the cavity through small holes arranged on a $(x,y)$ grid in the top plate. (b) Transmitted power $\left|S_{12}(\nu)\right|^2$ for different configurations ($\chi$) and for a given distance $d=\sqrt{x^2+y^2}$ between (1) and (2).\vspace{0.4cm}
    }
    \label{fig:setup}
\end{figure}

Thirty years after John's seminal work on the interplay between
photonic band gap formation and strong localization in disordered dielectric lattices~\cite{John1987}, a controlled experimental study of the optical transport properties in between ordered and disordered states of matter is still lacking~\cite{Sperling2016}.
Here, we present experimental
results obtained for a 2D system composed of high index dielectric cylinders in air~\cite{Laurent2007} placed according to SHU point patterns~\cite{Florescu2009}.
To probe the different transport regimes experimentally, we conduct measurements in the microwave regime since the frequency span in this regime is much larger than in the optical one. Furthermore, our microwave setup provides a more versatile platform compared to optics.
Our samples consist of about $N\simeq 200$ cylindrical scatterers (dielectric permittivity $\varepsilon\simeq 36$, radius $r=3$\,mm, height $h=5$\,mm; the Mie scattering efficiency of such a cylinder is shown in the Supplemental Material, Fig.~%
\ref{fig:scatteringMeanFreePath})
%S1)
placed in an aluminum 2D cavity ($50\times 50\times 0.5$\,cm$^3$) on a SHU point pattern (on a square of size of approximately $25\times 25$~cm$^2$) generated by simulating an annealing relaxation scheme~\cite{Froufe-Perez2016} (see Fig.~\ref{fig:setup}(a)).
We perform measurements on five different configurations $\chi=0.15, 0.25, 0.30, 0.40$, and a triangular lattice.
For all the samples studied, we kept the number density constant ($\rho\simeq 0.32$~cm$^{-2}$).
The point patterns and the structure factors of the samples are shown in the Supplemental Material Fig.~%
\ref{fig:Sk}.
%S2.
The cavity can be considered as two dimensional for the microwave frequencies $\nu<10$\,GHz studied. Under this condition, only the first transverse magnetic mode, TM$_0$, exists in air: the electric field is perpendicular to the plane, and the field amplitude is uniform over the cavity height~\cite{Jackson1998}.
We mimic an infinite 2D system by placing absorbing carbon loaded polyurethane foam between the sample and the metallic walls of the cavity.
We raster the cavity with a mobile antenna that is inserted by a robotic arm through holes drilled into the upper plate with a diameter 2\,mm, on a $5\times 5$\,mm$^2$ grid unit cell.
Considering the sample size, and the fact that we are not able to penetrate the cavity at the holes above the scatterers, we end up with about $\sim 2700$ measured positions.

At each grid point $(x,y)$, we measure the complex transmission spectrum $S_{12}(\nu)$ between a fixed antenna (1) placed at the center of the cavity and the mobile antenna (2) using a vector network analyzer.
Figure~\ref{fig:setup}(b) shows examples of measured spectra $|S_{12}(\nu)|^2$ between the central position $1$ and probe position $2$ for different $\chi$ values and for a given distance $d$ between the antennas. The small transmission values of order $10^{-6}$ or less are because the receiving antenna is weakly coupled  to the cavity.
The measured spectra consist of a superposition of peaks which are associated to the resonances of the system.
We extract their frequency, complex amplitude and width using harmonic inversion as described in Ref.~\cite{Main2000,Wiersig2008}.
We then cluster the resonances measured on all the lattice points in order to reveal all the eigenmodes present in the system without being spoiled by false resonances induced by noise (See Supplemental Material~%
\ref{sec:clusteringIntoModes}%
%§~IIIA%
~\cite{Maier1952,Pourrajabi2014,Ruiz2007}).

\begin{figure}
    \includegraphics[width=\columnwidth]{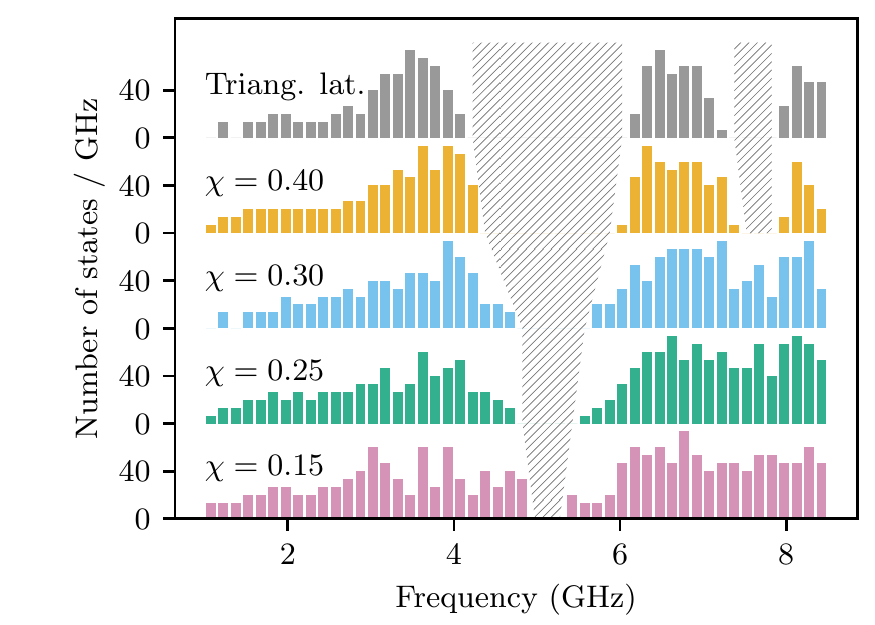}
    \caption{Experimental density of states (DOS). Histogram of states per $0.15$ GHz frequency interval for different configurations: $\chi$ between 0.15 and 0.40, and for the triangular lattice. The hatched areas are a guide to the eye to illustrate the measured band gap widths as a function of $\chi$.}
    \label{fig:NOS_hist}
\end{figure}
In Fig.~\ref{fig:NOS_hist}, we plot a histogram of the frequencies of the eigenmodes, which is directly proportional to the density of states (DOS). We compare the results for SHU point patterns with different values of $\chi$, to the results obtained for a triangular lattice.
As shown in earlier numerical work, the triangular lattice is the champion photonic crystal structure in 2D, with a gap slightly larger than disordered hyperuniform structures~\cite{Froufe-Perez2016}.
Our experimental data confirms the two first TM photonic band gaps predicted for the triangular lattice~\cite{Joannopoulos2008}.
We also find frequency windows without states for the SHU disordered systems.
Surprisingly, not only the first but also the second band gap is present in the $\chi=0.4$ sample.
To our knowledge, second and higher order band gaps have so far neither been predicted nor observed in disordered systems.
This finding is in contradiction to previous claims about the origin of band gaps in disordered photonic materials~\cite{Jin2001,Miyazaki2003,Rockstuhl2006}.
To corroborate additional evidence for this interesting observation, we performed band structure calculations, using the same parameters as in the experiment (see Supplemental Material §.~%
\ref{sec:numericalDOS}%
%IV%
~\cite{mpb}).
These numerical data confirm the existence of a second-order band gap for $\chi \ge 0.4$.
Both the first and the second gap approximately match the maxima of $S(k)$ of the triangular lattice
and of the SHU structures, supporting earlier proposals that short-range spatial correlations play a key role for the opening of band-gaps in amorphous photonic materials~\cite{Froufe-Perez2016}.
Experimentally, we observe a narrow photonic band gap even for our most disordered sample ($\chi=0.15$). Our numerical data for a large ensemble of system realizations, however, suggest that the band gap closes for $\chi \lesssim 0.3$ and reduces to a pseudogap with a small but finite density of states.
Naturally, variations between different realizations of hyperuniform materials become more pronounced for smaller values of $\chi$ (see Supplemental Material Fig.~%
\ref{fig:largestGapStats})
%S5)
and moreover the number of states per frequency bin is small for a finite sized system. This can lead to the situation that the central frequency and width of the band gaps depend on the precise realization of the point pattern, which is a distinct feature of disordered materials not found in crystals. For larger values of $\chi$ these variations are suppressed, and the gap becomes more robust against statistical fluctuations.

\begin{figure*}
    \centering
    \includegraphics[width=\textwidth]{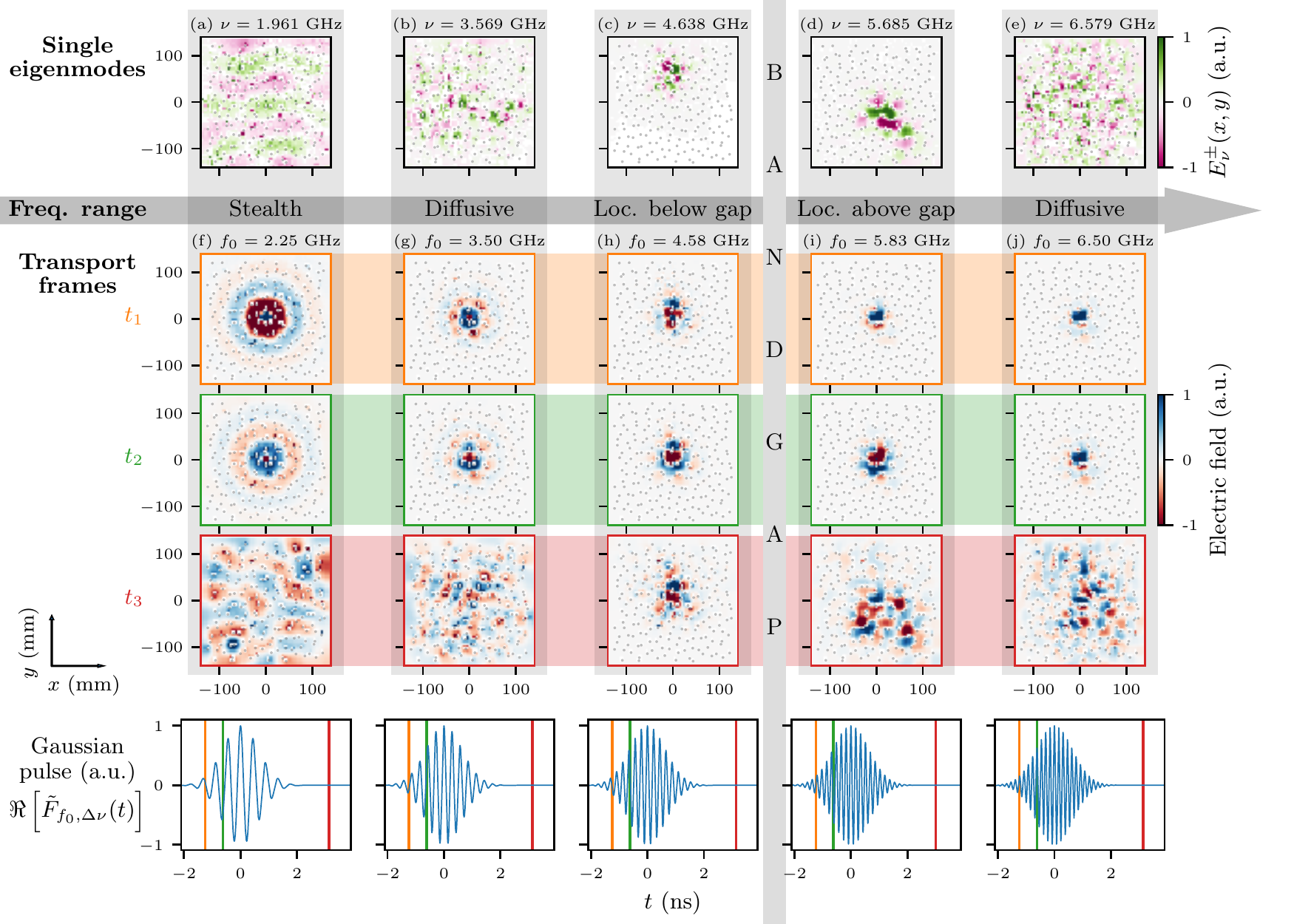}
        \caption{
       Electromagnetic field distribution of the eigenmodes and wave transport in the time domain for a sample with $\chi=0.30$ ($\nu_\mathrm{c}=2.88$GHz).
       (a-e): Signed amplitudes of selected eigenmodes at different characteristic frequencies.
       (a) cavity mode, (b) diffusive mode, (c) dielectric localized mode, (d) air localized mode and (e) diffusive mode. (f-j): Maps of the electric field for wave transport at different times $t_1,t_2,t_3$ and for different central frequencies $f_0$.
       The wave---a Gaussian pulse centered at $f_0$ and having a width of 0.5\ GHz in the frequency domain---is emitted at the center of the maps, and its temporal representation is shown in the last line ($\Re [\tilde F_{f_0,\Delta\nu}(t)]$ is the real part of the Fourier transform of the Gaussian band pass filter).
       The colored vertical lines indicate the time of each frame shown $t_1,t_2,t_3$. Entire videos are included in the Supplemental Material, Videos~%
       \ref{fig:videos}.
       %S6.
       The color scale is adjusted for each individual panels.
       }
    \label{fig:singleModesTransport}
\end{figure*}

We now consider the optical properties of our material outside the gap~\cite{Froufe-Perez2017}.
The amplitude of the peaks observed in Fig.~\ref{fig:setup}(b), and clustered to reveal the eigenmodes, differs from one position to the other and from this we obtain an electric field amplitude map $E_\nu(x,y)$ of an eigenmode~\cite{Stein1995} (see Supplemental Material §~%
%III%
\ref{sec:clustering}%
~\cite{Maier1952,Pourrajabi2014,Ruiz2007,Xeridat2009}).
These eigenmodes maps, shown in the first line of Fig.~\ref{fig:singleModesTransport}, reveal the striking variations in optical transport properties across the spectral range covered by our experiment.
At low frequencies, we observe simple square cavity modes as if the medium was homogeneous, which is a remarkable result given the fact that at $\nu\sim 2$GHz, the system size $L=25$\ cm is almost two orders of magnitude larger than the Boltzmann mean free path $\ell_\mathrm{s}(\nu)$ of the cylinder ensemble (see Supplemental Material Fig.~%
\ref{fig:scatteringMeanFreePath}),
%S1),
with $\ell_\mathrm{s}(\nu)=[\sigma_\mathrm{s}(\nu)\rho]^{-1}$ given by the total scattering cross section $\sigma_\mathrm{s}(\nu)$ and the number density $\rho$.
An alternative way to study wave propagation in the SHU material is to monitor the wave emitted by the central antenna as it propagates through the medium in the time domain. By calculating the real part of the Fourier-transform of $S_{12}(\nu)\times F_{f_0,\Delta\nu}(\nu)$ (with $F_{f_0,\Delta \nu}$ a band pass filter of bandwidth $\Delta\nu$ centered around $f_0$) at all points on the lattice, we reconstruct movies of the propagating electromagnetic fields as a function of time for the selected bandwidth $\Delta \nu$.
Individual frames of the movies are shown in Figs.~\ref{fig:singleModesTransport}(f-j) (details on the numerical procedure and the entire movies are included in the Supplemental Material §~%
\ref{sec:videos}).
%V).
Figure~\ref{fig:singleModesTransport}(f) shows that at low frequencies a circular wave propagates from the central antenna into the medium again signaling transparency. 
Note that the disordered pattern observed at $t_3$ in Fig.~\ref{fig:singleModesTransport}(f) is due to the nonperfectly absorbing foams placed around the sample which reflect part of the signal (for more details, see Supplemental Material Videos~%
\ref{fig:videos}-\ref{vid:stealth1.5} and \ref{fig:videos}-\ref{vid:stealth2.0}).
%S6-1,2).
From the velocity of the circular wave in the medium we can derive the effective refractive index of the samples and find $n_\mathrm{eff} \sim 1.8$.
Equally, counting the nodal lines of the modes (Fig.~\ref{fig:singleModesTransport}(a)) and relating them to their frequencies, we obtain values of the effective refractive index of the metamaterial in the range $n_\mathrm{eff}=1.7\pm 0.3$. The uncertainty is due to the fact that, for disordered systems, the cavity size is not well defined and moreover, we observe a slight increase of $n_\mathrm{eff}$ from $\nu=1 \to 3$ GHz. For comparison, the Maxwell-Garnett effective refractive index, which in 2D corresponds to the square root of the surface averaged permittivity, is $n_\text{MG}=2.05$. 

Torquato and coworkers named their designer materials ``stealthy'' hyperuniform because they predicted them to be fully transparent below a threshold frequency $\nu<\nu_\mathrm{c}$~\cite{Batten2008}.
The latter is equivalent to saying that $L/\ell^\star \to 0$ (with $\ell^\star$ the transport mean free path), while $L/\ell_\mathrm{s}$ remains finite and can even be larger than one.
In this first-order or single-scattering approximation $\nu_\mathrm{c}=\frac{c}{n_\mathrm{eff}}\sqrt{\frac{\rho\chi}{\pi}}$~\cite{Froufe-Perez2017}. For our system parameters, the theoretical $\nu_\mathrm{c}$ range from $\simeq 2.2$\ GHz ($\chi=0.15$) to $\simeq 3.0$\ GHz ($\chi=0.4$) based on an effective refractive index of $n_\mathrm{eff} \sim 1.8$.
\citet{Leseur2016} demonstrated recently that stealthy transparency is also robust against recurrent multiple scattering.
They establish a stricter criterion for transparency, $L/\ell_\mathrm{s}\ll k \ell_\mathrm{s}$, in a dense SHU disordered material composed of dipolar point scatterers. While transparency is retained under this condition it also implies that the transition at $\nu_\mathrm{c}$ is not sharp but system size dependent. 
From a theoretical evaluation of $\sigma_\mathrm{s}(\nu)$ for our $\varepsilon=36$ cylinders in air, however, we find that only for $\nu < 1$ GHz the condition $L/\ell_\mathrm{s} < k \ell_\mathrm{s}$ is met (see Supplemental Material Fig.~%
\ref{fig:scatteringMeanFreePath}~\cite{Bohren1998}%
%S1~\cite{Bohren1998}%
).
The experimental results, however, suggest that the condition set by \citet{Leseur2016} is too restrictive and transparency remains a robust feature for $\nu<\nu_\mathrm{c}$ in our dense, high index SHU materials, even for $k \ell_\mathrm{s}\lesssim 1$ (see also Supplemental Material Fig.~%
\ref{fig:stealthDiffusive}).
%S7).

For frequencies $\nu>\nu_\mathrm{c}$ transparency is clearly lost and we observe scattering and wave diffusion. The modes become disordered, Fig.~\ref{fig:singleModesTransport}(b), and the propagating wavefronts in the time domain are highly distorted signaling mean free paths smaller than the system size, Fig.~\ref{fig:singleModesTransport}(g).
A closer inspection of the propagating wave fronts, Supplemental Material Fig.~%
\ref{fig:stealthDiffusive},
%S7,
illustrates how the onset of scattering and wave diffusion is shifted to higher frequencies $\nu_\mathrm{c}(\chi)\propto \sqrt{\chi}$ as the system becomes more and more stealthy.
At frequencies close to the first band gap, we observe spatially localized modes as shown in Figs.~\ref{fig:singleModesTransport}(c) and (d)~\cite{Laurent2007,LeThomas2009,Garcia2012}.
In the time domain, we find that, at longer times, the wave stays localized near the central antenna, as shown in the panels framed red in Figs.~\ref{fig:singleModesTransport}(h,i) and in the corresponding Supplemental Material videos~%
\ref{fig:videos}-\ref{vid:beforeBG} and \ref{fig:videos}-\ref{vid:afterBG}. 
%S6-4 and S6-6.
\begin{figure}
       \includegraphics[width=\columnwidth]{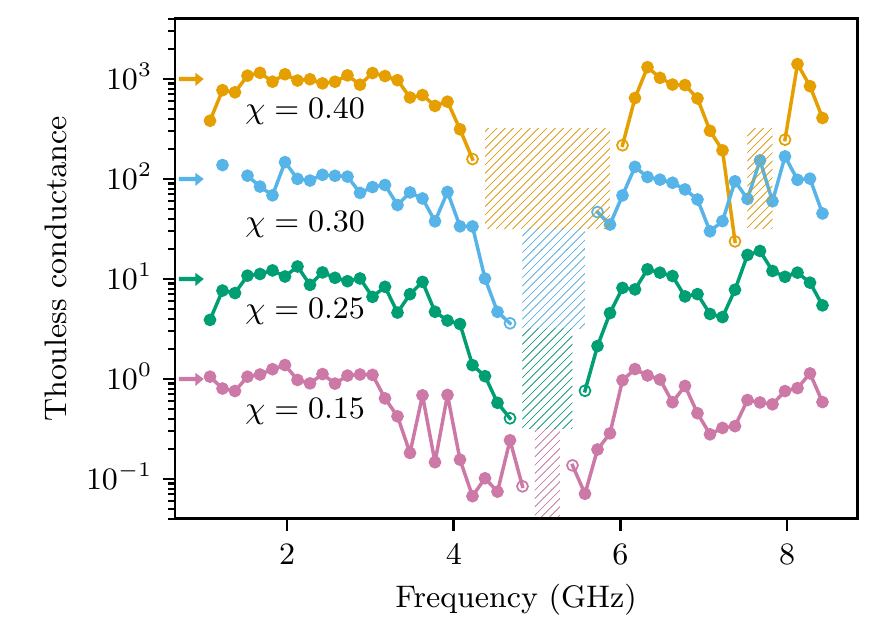}
    \caption{Thouless conductance for different degrees of stealthy hyperuniformity $\chi$ between 0.15 and 0.40. The curves are shifted by a factor 10 for clarity. The hatched areas show the width of the experimentally observed band gaps for each value of $\chi$ using the same colors.
    }
    \label{fig:thouless_hist}
\end{figure}
We note that the modes below the band gap are localized on the dielectric cylinders, Fig.~\ref{fig:singleModesTransport}(c), and the modes above the band gap are localized in air, Fig.~\ref{fig:singleModesTransport}(d).
For frequencies in between the first and the second band gap we again observe diffusive modes, Fig.~\ref{fig:singleModesTransport}(e), as well as extended waves at later times, Fig.~\ref{fig:singleModesTransport}(j). For frequencies in the band gaps we find no modes, all positions are phase coherent and there is no propagation. 

Next, we calculate the Thouless conductance $g_\mathrm{Th}= \delta \nu/\Delta \nu$, which is a fundamental localization parameter~\cite{Thouless1977,Wang2011,Mondal2019}.
Thouless argued that in the Anderson localization regime, the dimensionless ratio $g_\mathrm{Th}= \delta \nu/\Delta \nu$ falls below unity. 
In this case, the spectral widths $\delta \nu$ of the
modes are smaller than their spacing $\Delta \nu$, and the modes are isolated ~\cite{Thouless1977}. In the opposite limit, for $g_\mathrm{Th}\geq 1$ modes overlap and waves can propagate. 
By calculating the average width of the modes in each frequency bin, Fig.~\ref{fig:NOS_hist}, we extract the mean Thouless conductance for each frequency bin as shown in Fig.~\ref{fig:thouless_hist}.
We have marked the data points directly at the band edges by open circles in Fig.~\ref{fig:thouless_hist}. 
Note that, due to the discretization, their values can be affected by the zeroes of the DOS in the gap. Inside the band gap there are no modes and $\left<g_\mathrm{Th}\right>$ is not defined.
We find values of $\left<g_\mathrm{Th}\right> \sim 1$ everywhere except in the vicinity of the gap where $\left<g_\mathrm{Th}\right>$ drops by up to two orders of magnitude, signaling localization.
This result is consistent with both the finite spatial extension of the modes we observe experimentally, see Figs.~\ref{fig:singleModesTransport}(c,d), and the localization of the propagating wave in the same frequency domain, Fig.~\ref{fig:singleModesTransport}(h,i).
In the low-frequency regime, the Thouless conductance is close to one, and wave transport expands over the whole system size. 

In conclusion, we show experimentally that disordered dielectric structures display different characteristic transport regimes such as transparency, photon diffusion, Anderson localization, as well as first and even second order band gaps. We rationalize our findings by analyzing the mode structure and the propagation of waves in the time domain. We find evidence that transparency is robust against recurrent multiple scattering, and that the stealthy materials we study retain their low-frequency transparency even for the unusually strong refractive index mismatch between our scatterers and air $\sqrt{\varepsilon/\varepsilon_\text{air}}=6$. Our results lend support to recent numerical predictions and shed new light on the interplay between disorder and correlations~\cite{Froufe-Perez2017}. We believe this will have significant consequences for the design of photonic materials, such as two-dimensional nanostructured materials for light harvesting in solar cells~\cite{Vynck2012} or light guiding in all-optical circuit applications~\cite{Milosevic2019}.

\subsection*{Acknowledgments}
G.A., L.S.F., and F.S. acknowledge funding by the Swiss National Science Foundation through Project No. 169074 and No. 188494, and through the National Center of Competence in Research Bio-Inspired Materials. We would like to thank Paul Chaikin and Juanjo Saenz for discussions.

\bibliography{sharedBiblio}

\clearpage

%% the following solution has the draw back that the link of Fig. S2 points on Fig. 2.
%\setcounter{figure}{0}
%renewcommand{\thefigure}{S\arabic{figure}}
%% better solution
\setcounter{maintextfigures}{\value{figure}}%we store the number of figures in the main text
\renewcommand{\thefigure}{S\the\numexpr\value{figure}-\value{maintextfigures}\relax}
\setcounter{equation}{0}
\renewcommand{\theequation}{S\arabic{equation}}

\setcounter{secnumdepth}{3}

\onecolumngrid

\vspace*{-0.2cm}
\section*{Supplementary Material}

This document contains the scattering properties of a single rod, details on
the structures of the point patterns, the band structure calculation,
details on the time domain propagation videos and all the technical
information on the data analysis.
The seven videos (permanently stored on the Zenodo repository: \url{https://doi.org/10.5281/zenodo.3978032})
show how the electromagnetic wave propagates in the
cavity for different frequency ranges (see Supplemental Material Fig.~\ref{fig:videos}
for the description of the videos).

\vspace{1cm}

\twocolumngrid

\section{Boltzmann scattering mean free path.}
\begin{figure}[b]
    \centering
    \includegraphics[width=\columnwidth]{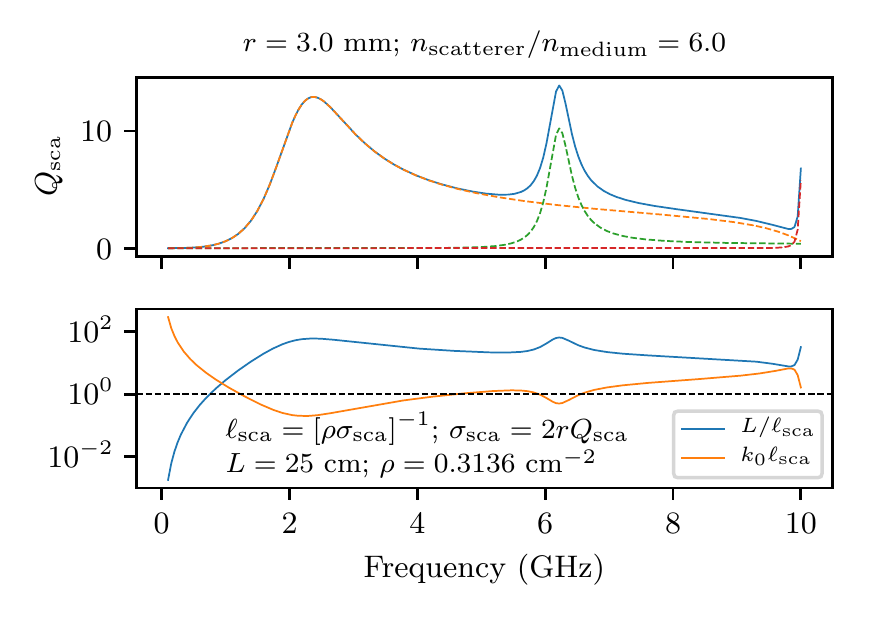}
    \caption{Upper panel: Scattering efficiency $Q_\mathrm{sca}$ of individual cylinders in TM polarization (solid blue line), and the three first terms in the Mie expansion (dashed lines). Lower panel: optical density $L/\ell_\mathrm{s}$ in the independent scattering approximation using the Boltzmann scattering mean free path and the sample size $L$. Also shown is $k_0 \ell_\mathrm{s}$ with $k_0=2\pi/\lambda$ and the wavelength in vacuum $\lambda_0$ ($k_0=2\pi/\lambda_0$).}
    \label{fig:scatteringMeanFreePath}
\end{figure}

In Fig.~\ref{fig:scatteringMeanFreePath} we show the scattering efficiency $Q$ of an individual cylinder in TM polarization calculated using Mie theory~\cite{Bohren1998} (upper panel).
In the lower panel, we show how the corresponding Boltzmann scattering mean free path $\ell_\mathrm{sca}(\nu)=[\sigma_\mathrm{sca}(\nu)\rho]^{-1}$ (with $\sigma_\mathrm{sca}(\nu)=2 r Q_\mathrm{sca}$ the total scattering cross section) compares with $L$, the size of the system, and $\lambda_0$, the wavelength in vacuum of the wave.

\newpage

\section{Point patterns and their structure factors.}

\begin{figure}[b]
    \centering
    \includegraphics[width=\columnwidth]{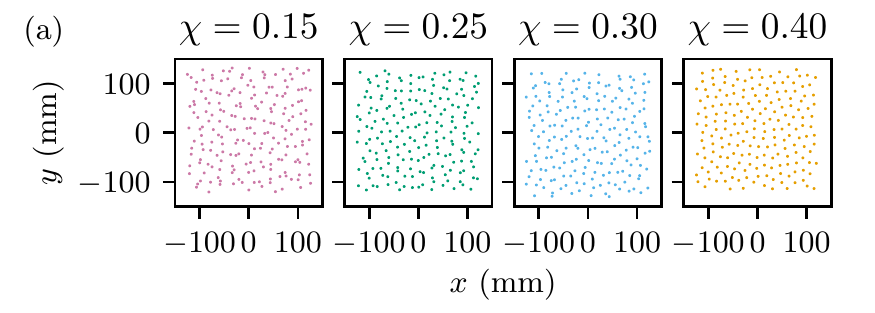}
    \includegraphics[width=\columnwidth]{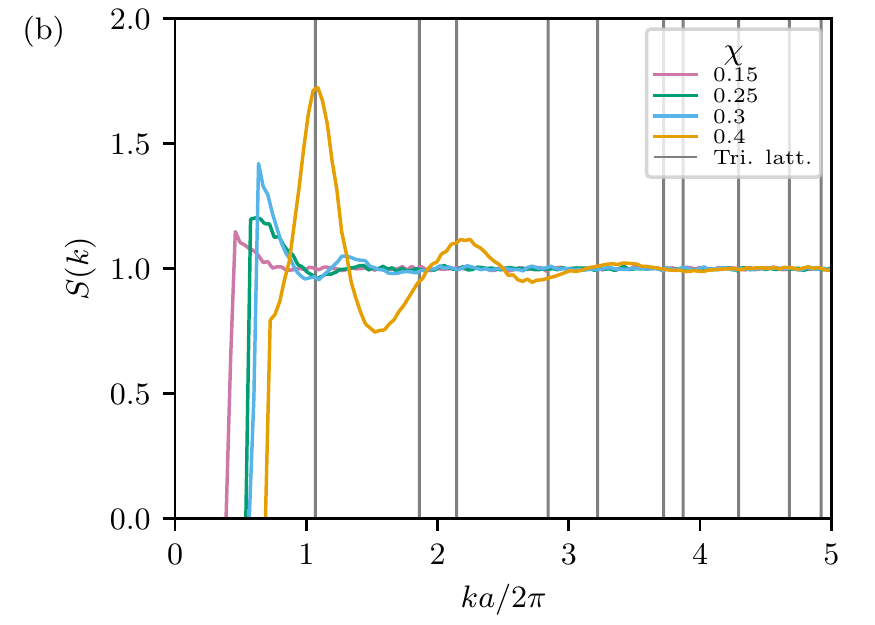}
    \caption{(a) Point patterns of the studied samples. (b) Radially averaged structure factors $S(k)$ of the studied samples as a function of $ka$, where $a=1/\sqrt{\rho}$ and $\rho$ denotes the number density of scatterers. The structure factors are averaged over 1000 different realizations of about 200 points. The grey vertical lines indicate the peaks of the radially averaged triangular lattice structure factor (Bragg peaks).}
    \label{fig:Sk}
\end{figure}

Figure~\ref{fig:Sk}(a) shows the point patterns of the samples studied in this study, and Fig.~\ref{fig:Sk}(b) the corresponding average structure factors
\begin{align}
S(\mathbf {k} )={\frac {1}{N}}\sum _{j=1}^{N}\sum _{l=1}^{N}\mathrm {e} ^{-i\mathbf {k} \cdot (\mathbf {R} _{j}-\mathbf {R} _{l})},
\end{align}
over 1000 samples generated as the ones used in this study,
where $R_j$ are the positions of the $N$ points, and $\mathbf{k}$ is the wavevector.

\clearpage

\section{Visualization of the eigenmodes of the disordered cavity}
\label{sec:clustering}

\subsection{Clustering of the resonances into modes}
\label{sec:clusteringIntoModes}

\begin{figure}
    \centering
    \includegraphics[width=\columnwidth]{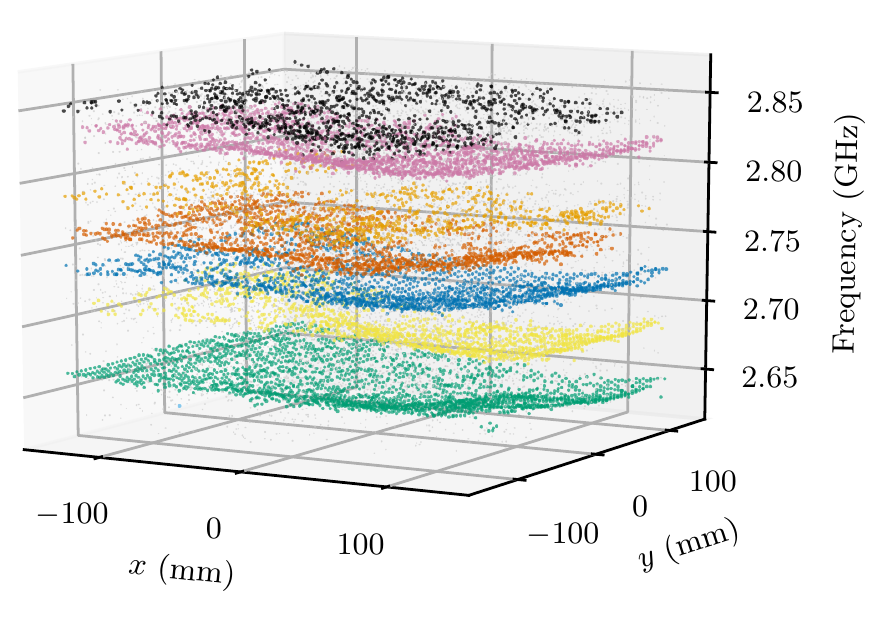}
    \includegraphics[width=\columnwidth]{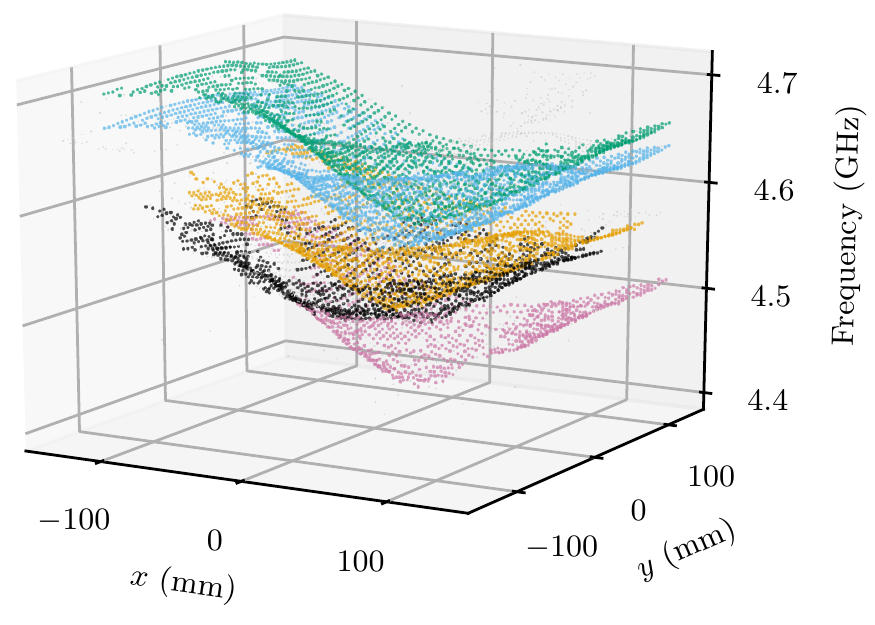}
    \caption{For each position $(x,y)$, a spectrum is measured and the frequencies are extracted using harmonic inversion: these are the points plotted in this figure for two different frequency ranges.
    The points are then clusterized: each color corresponds to a cluster found by the algorithm.
    The upper panel corresponds to a typical situation in the stealth regime where the intensity is almost uniform over the sample (small frequency shifts).
    The lower panel corresponds to the case of localized modes with large intensities corresponding to large frequency shifts.}
    \label{fig:clustering}
\end{figure}

The measured spectra consist of a superposition of peaks (see Main Text Fig.~%
%1(b))
\ref{fig:setup}(b)).
which are associated to the resonances of the system.
We determine the frequencies $\nu^i$, widths $\gamma^i$ and complex amplitudes $A^i$ of each resonance $i=1,\dots, N$
using the harmonic inversion method described in ref.~\cite{Main2000,Wiersig2008}.
Ideally, resonances belonging to the same mode should all have the same frequency.
In practice, the presence of the mobile antenna at every point $(x,y)$ shifts the resonant frequency by a small amount depending on the intensity of the electromagnetic field at the specific mobile antenna position~\cite{Maier1952}, see Fig.~\ref{fig:clustering}.
Note that we minimize this perturbation due to the mobile antenna by having it extending into the cavity by only 1\ mm whereas the height of the cavity is 5\ mm.
This has the consequence that it is weakly coupled to the field, and explains the low transmission values as seen in Main Text Fig.~%
\ref{fig:setup}(b).
%1(b).
We identify all data points belonging to a certain cluster by using a density-based clustering algorithm~\cite{Ruiz2007} fulfilling the condition that two points having the same coordinate $(x,y)$ cannot be in the same cluster.
To associate each resonant signal at position $(x,y)$ to a specific mode, we apply a semi-supervised clustering algorithm.
This allows us to identify every single mode of the disordered cavity, associated with discrete resonance frequencies, as long as the mode amplitude is large enough to be detected by the vector network analyzer~\cite{Pourrajabi2014,Ruiz2007}.

More precisely, we use a slightly modified version of the C-DBSCAN algorithm published in Ref.~\cite{Ruiz2007}.
In our version, step 2 of the algorithm~\cite{Ruiz2007} either labels the points in the KD-tree leaf as noise ratio (if the density is too small), or we create a local cluster for each point in the leaf. Depending on the frequency range, we run our modified version of C-DBSCAN either in the $(x,y,\nu)$, $(x,y,\nu,\gamma)$ or $(x,y,\nu,\gamma,\ln A)$ space to reach the best clustering results.
An example of the result is shown in Fig.~\ref{fig:clustering} where the different clusters, or modes, found by the algorithm are plotted using different colors.

\subsection{Electric field amplitude maps}
\label{sec:signedAmplitude}

In the first line of Main Text Fig.~%
\ref{fig:singleModesTransport},
%3,
we plot the signed amplitude $E_\nu^\pm(x,y)=\sgn\left(\textrm{Re}[\tilde{S}_{12}]\right)\vert \tilde{S}_{12}\vert$, where $\tilde{S}_{12}$ is the transmission deduced from $S_{12}$ after the ad hoc rotation of the global phase making the real and imaginary parts statistically independent~\cite{Xeridat2009}.
This allows to represent both the real and imaginary parts of the eigenmodes on the same map.

\clearpage
\section{Numerical simulations of the DOS}
\label{sec:numericalDOS}
\begin{figure}
    \centering
    \includegraphics[width=\columnwidth]{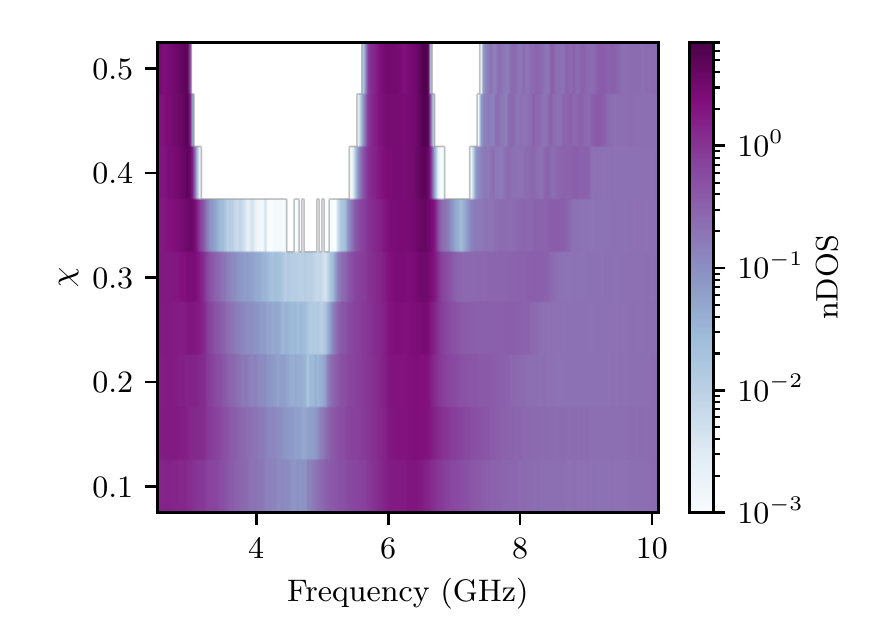}
    \caption{Normalized density of states (nDOS) obtained by taking the average over the band structure calculated numerically for 500 system realizations at each value of $\chi$.}
    \label{fig:numericalDOS}
\end{figure}

Figure~\ref{fig:numericalDOS} shows the normalized density of states (nDOS) of the stealthy hyperuniform samples obtained numerically for a large statistical ensemble of point patterns and using periodic boundary conditions.
The properties of the dielectric cylinders and their density are identical to those of the system studied in the experiment.
The nDOS was calculated using the MIT Photonic Bands~\cite{mpb} software using the supercell method~\cite{Joannopoulos2008} as described earlier in ref.~\cite{Froufe-Perez2016}.
This dataset was obtained by calculating 500 different samples for each $\chi$-value (between 0.1 and 0.5, every 0.05).

\begin{figure}
    \centering
    \includegraphics[width=\columnwidth]{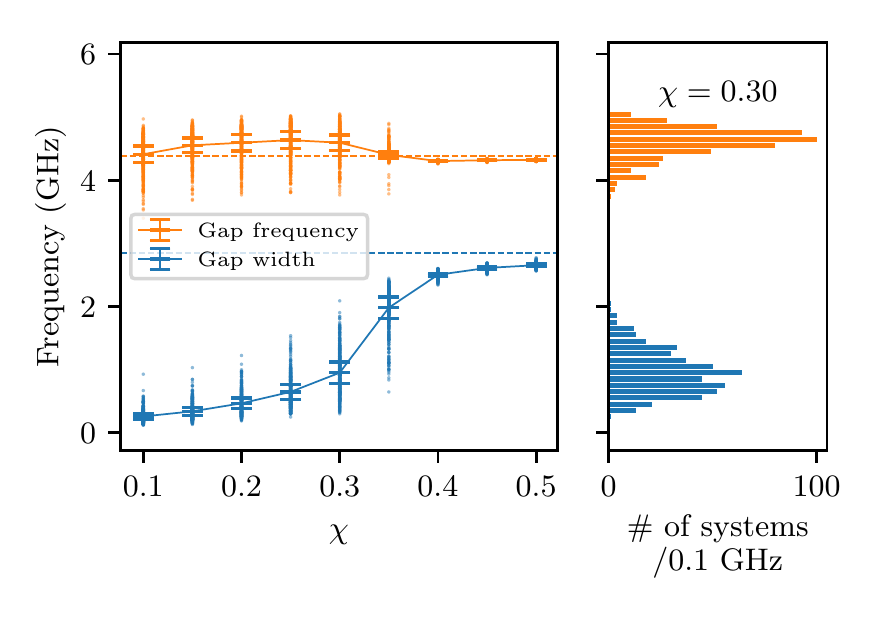}
    \caption{Spread of the first gap central frequency and width found in the numerical results used to obtain Fig.~\ref{fig:numericalDOS}. The error bars correspond to the standard deviations, the scattered points to the 500 individual systems per $\chi$-value used to compute the statistics.
    The dashed lines correspond to the results obtained for the triangular lattice.
    The right panel shows the histograms for the $\chi=0.30$ samples.
    }
    \label{fig:largestGapStats}
\end{figure}

Figure~\ref{fig:largestGapStats} shows the average and the standard deviation of the gap central frequency and width found for the samples used in Fig.~\ref{fig:numericalDOS}.
The statistical variations are large at low and intermediate $\chi$-values (between 0.10 and 0.35).
At large $\chi$-values ($\geq 0.4$), the standard deviation vanishes: the gap central frequencies and widths are similar from sample to sample.
%\vfill\null
%\pagebreak
%\columnbreak

%\clearpage

\section{Time domain propagation videos}
\label{sec:videos}

We obtain time domain propagation signals from the real part of the Fourier transform of the complex transmission spectra multiplied by a chosen bandpass filter centered at $f_0$ with a standard deviation $\Delta \nu$. We use a Gaussian bandpass filter to avoid window effects in the Fourier transform. The excitation in the time domain is therefore a Gaussian pulse with a temporal spread inversely proportional to $1/\Delta \nu$ of the Gaussian bandpass filter.

\begin{figure*}
    \centering
    \includegraphics[width=\textwidth]{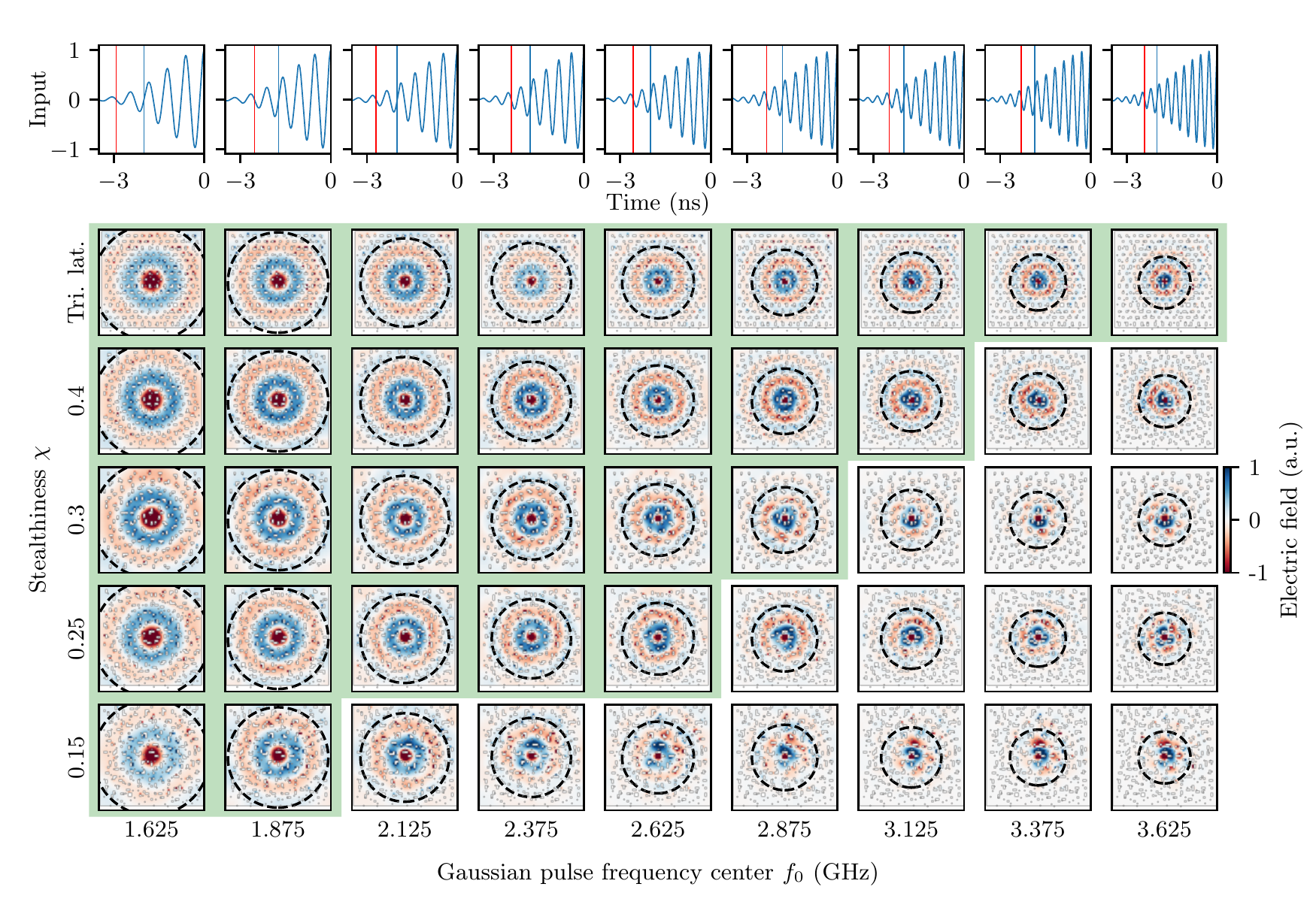}
    \caption{Maps of the electric field amplitude for the propagation of a pulse of spectral width $\Delta\nu=0.125$\ GHz at different central frequencies $f_0$ (for details see text and Main Text Fig.~%
 \ref{fig:singleModesTransport}),
 %3,
 and first half of the Gaussian pulse used for the excitation.
    The frames shown in the figure are taken at the time marked by the blue vertical line.
    The panels in the green polygon indicate frequencies below $\nu_\mathrm{c}(\chi)$.
    The radius of the dashed circles indicate the place where a wave emitted at the time marked by the red vertical line should be at the time marked by the blue vertical line, for a homogeneous medium with $n_\mathrm{eff}=1.8$.
    The color scale is adjusted for each individual panels.
    }
    \label{fig:stealthDiffusive}
\end{figure*}

Videos~\ref{fig:videos}-\ref{vid:stealth1.5}, \ref{vid:stealth2.0} and \ref{vid:diffLF} show the propagation of the wave in the low frequency  regime (well below the gap frequency $\nu_\mathrm{G}\simeq 5$\ GHz.
\begin{figure*}
    \begin{enumerate}
        \item \label{vid:stealth1.5} Stealth regime (Gaussian bandpass filter, $f_0=1.75$\ GHz, $\Delta \nu=0.25$\ GHz)
        \item \label{vid:stealth2.0} Stealth regime (Gaussian bandpass filter, $f_0=2.25$\ GHz, $\Delta \nu=0.25$\ GHz)
        \item \label{vid:diffLF} Wave diffusion (Gaussian bandpass filter, $f_0=3.5$\ GHz, $\Delta \nu=0.25$\ GHz)
        \item \label{vid:beforeBG} Dielectric Anderson localized modes just below the band gap (Gaussian bandpass filter, $\Delta \nu= 0.25$~GHz)
        \item \label{vid:bandgap} Square filter in the band gaps
        \item \label{vid:afterBG} Air Anderson localized modes just above the band gap (Gaussian bandpass filter, $\Delta \nu = 0.25$~GHz)
        \item \label{vid:diffHF} Wave diffusion (Gaussian bandpass filter, $f_0=6.5$\ GHz, $\Delta \nu=0.25$\ GHz)
    \end{enumerate}
    \caption{Videos description. The videos are permanently stored on the Zenodo repository: \url{https://doi.org/10.5281/zenodo.3978032}.}
    \label{fig:videos}
\end{figure*}
We observe that for frequencies $\nu<\nu_\mathrm{c}$ and at early times, the spherical wave structure is well preserved, indicating the absence of scattering.
This boundary between the stealth regime and the diffusive regime is also shown in more detail in Fig.~\ref{fig:stealthDiffusive}.
The panels in the green shaded polygon indicate that the Gaussian pulse central frequency $f_0$ is below the critical stealth frequency $\nu_\mathrm{c}=\frac{c}{n_\mathrm{eff}}\sqrt{\frac{\rho\chi}{\pi}}$, and above $\nu_\mathrm{c}$ elsewhere.
By eye, we see a clear correlation between the wave front smoothness and the transition from the stealth regime to the diffusive regime for frequencies $\nu>\nu_\mathrm{c}$.
Since $\nu_\mathrm{c}\propto \sqrt{\chi}$ the transition is shifted to higher frequencies when increasing the degree of stealthiness $\chi$.
Note that the wave distortion at later times (in the videos) is explained by reflections of the wave on the non-ideal absorbing foam walls.

Video~\ref{fig:videos}-\ref{vid:beforeBG} (respectively \ref{fig:videos}-\ref{vid:afterBG}) shows the electromagnetic field for a Gaussian pulse centered 0.25\ GHz below (resp. above) the band gap and having a width $\Delta\nu=0.25$\ GHz.
Video~\ref{fig:videos}-\ref{vid:diffHF} shows the propagation of the wave in the high frequency regime, well above the first band gap.
As in the low frequency regime for frequencies above $\nu_\mathrm{c}$, we observe a strong scattering and wave diffusion.

Finally, video~\ref{fig:videos}-\ref{vid:bandgap} shows the electromagnetic field in the band gap. For this video, the bandpass filter was chosen to be a square filter fitting exactly the band gaps as extracted from Main Text Fig.~%
\ref{fig:NOS_hist}.
%2.
This explains the windowing effect seen in the input signal.

\end{document}